\documentstyle[12pt]{article}
\newcommand{\cz}{C_0}
\newcommand{\cuu}{C_{11}}
\newcommand{\cud}{C_{12}}
\newcommand{\mg}{m_{\tilde{g}}}
\newcommand{\mh}{M_{H^{+}}}
\newcommand{\mhs}{M^2_{H^{+}}}
\newcommand{\beq}{\begin{equation}}
\newcommand{\eeq}{\end{equation}}
\newcommand{\beqn}{\begin{eqnarray}}
\newcommand{\eeqn}{\end{eqnarray}}
\newcommand{\stackM}{\stackrel{\scriptstyle >}{{ }_{\sim}}}
\newcommand{\stackm}{\stackrel{\scriptstyle <}{{ }_{\sim}}}  
\begin{document}

\thispagestyle{empty}
\def\pubnum{376}
\def\data{November, 1995}
\begin{flushright}
{\parbox{3.5cm}{
UAB-FT-376

November, 1995

hep-ph/9511292
}}
\end{flushright}
\vspace{3cm}
\begin{center}
\begin{large}
\begin{bf}
SUPERSYMMETRIC QCD CORRECTIONS TO THE TOP QUARK DECAY OF A HEAVY
CHARGED HIGGS BOSON\\
\end{bf}
\end{large}
\vspace{1cm}
Ricardo A. JIM\'ENEZ, Joan SOL\`A\\

\vspace{0.25cm} 
Grup de F\'{\i}sica Te\`orica\\ 
and\\ 
Institut de F\'\i sica d'Altes Energies\\ 
\vspace{0.25cm} 
Universitat Aut\`onoma de Barcelona\\
08193 Bellaterra (Barcelona), Catalonia, Spain\\
\end{center}
\vspace{0.3cm}
\hyphenation{super-symme-tric}
\hyphenation{com-pe-ti-ti-ve}
\begin{center}
{\bf ABSTRACT}
\end{center}
\begin{quotation}
\noindent
The supersymmetric QCD corrections to the hadronic width of a heavy charged
Higgs boson, basically dominated by the top-quark decay mode $H^{+}\rightarrow
t\,\bar{b}$, are evaluated at ${\cal O}(\alpha_s)$ within the MSSM and
compared with the standard QCD corrections. The study of such quantum effects,
which turn out to be rather large, is essential to understand the hypothetical
supersymmetric nature of a heavy charged Higgs boson potentially produced in
the near future at the Tevatron and/or at the LHC.
\end{quotation}
  
\newpage

\baselineskip=6.5mm  %(FOR PREPRINT)

After the discovery of the top quark at the Tevatron\,\cite{Tevatron}, the
``next-to-leading'' task in order of importance in Particle Physics is the
search for Higgs bosons. Indeed, whereas the finding of the top quark
completes the spectrum of matter fields in the Standard Model (SM) and it was
a necessary condition for the SM to be correct, it is not at all a sufficient
condition. The decisive proof lies in the nature of the spontaneous symmetry
breaking mechanism (SSB). Thus the very crucial question is still left
pending: is the SSB caused by truly fundamental scalars or is it triggered by
a dynamical mechanism involving a new species of fermions and/or a new strong
interaction force?. This question is as old as the proposal for the first
serious extensions and alternatives to the SM, namely, Supersymmetry (SUSY)
and Technicolour. Here we will be concerned with the Minimal Supersymmetric
Standard Model (MSSM)\,\cite{MSSM}.

In the near and middle future, with the upgrade of the Tevatron, the advent of
the LHC, and the possible construction of an $e^+\,e^-$ supercollider, new
results on top quark physics, and perhaps also on Higgs physics, will be
obtained in interplay with Supersymmetry that may be extremely helpful to
complement the precious information already collected at LEP from $Z$-boson
physics.  Here we wish to dwell on the phenomenology of supersymmetric Higgs
boson decays into hadrons with an eye on these future developments.  Two basic
parameters are needed to deal with the (tree-level) MSSM Higgs sector,
namely\,\, $(M_{A^0},\tan\beta$)\,\cite{Hunter}.  In the present study (see
also the companion paper\, \cite{CJS}), it will be useful to divide the
supersymmetric Higgs parameter space into two mass segments: light CP-odd
higgses ($M_{A^0}<M_Z$) and heavy CP-odd higgses ($M_{A^0}>M_Z$). However,
since these segments are unequally sensitive to large and small values of
$\tan\beta$, it will prove convenient to define the following two relevant
regions of the $(M_{A^0},\tan\beta)$-parameter space. One of them, call it
Region I, is characterized by Higgs masses in the light segment together with
large values of $\tan\beta$ of order $\sim m_t/m_b$. The other region (Region
II) consists of Higgs masses in the heavy segment in combination with more
moderate values of $\tan\beta$, say in the interval $2\stackm\tan\beta\stackm
30$\,\footnote{ Incidentally, we note that Regions I and II correspond
(approximately) to domains of the $(M_{A^0},\tan\beta$)-plane formerly
used\,\cite{GJS2} to alleviate some of the --still not completely banished--
$Z$-boson anomalies, where in addition it was required the concurrence of a
chargino and a stop both as light as permitted by phenomenological bounds,
namely, of about $60-70\,GeV$.}.
 
In Refs.\,\cite{GJSH,GJS}, whose notation and definitions we shall adopt
hereafter, we studied the impact of Region I on the top-quark decay process.
Here we will focus on a related process, namely, the top quark decay mode of a
charged supersymmetric Higgs, $H^{+}\rightarrow t\,\bar{b}$, which is allowed
in Region II, provided that $M_{A^0}\stackM 160\,GeV$.  We shall again be
partially concerned with Regions I and II in the companion paper\,\cite{CJS},
where we round off our study by addressing the quark-antiquark decay modes of
the three neutral Higgs states of the MSSM.

While a simple tree-level study of $H^+\rightarrow t\,\bar{b}$ is blind to the
nature of the Higgs sector to which $H^+$ belongs, it is clear that a careful
study of the quantum effects on $H^+\rightarrow t\,\bar{b}$ could be the clue
to unravel the potential supersymmetric nature of the charged Higgs; in
particular, it should be useful to distinguish it from a charged Higgs
belonging to a general two-Higgs doublet model.  The charged Higgs boson can
decay hadronically into several quark final states, and if it is sufficiently
heavy it can also decay into top and bottom quarks.  It is natural to tackle
the radiative effects on $H^+\rightarrow t\,\bar{b}$ by first considering the
QCD corrections.  The conventional QCD corrections were first considered in
Ref.\cite{LiOakes} and they are known to be large and negative in the limit of
Higgs masses much bigger than the quark masses.  In spite of the huge size of
the standard QCD corrections for light quark final states, it is clear that
these channels are severely suppressed by the small Yukawa couplings
$g\,m_q/M_W\ll 1 ({\rm for}\, q=u,d,c,s)$ and/or by off-diagonal CKM matrix
elements, so that their branching fractions are very tiny. Therefore, as soon
as the $t\,\bar{b}$ threshold is open ($M_{H^+}>m_t+m_b\sim 180\,GeV$) one is
left with $H^+\rightarrow t\,\bar{b}$ as the only relevant hadronic decay of a
heavy charged Higgs boson.  The conventional QCD corrections to that
decay\,\cite{LiOakes,Gambino} cannot distinguish the nature of the underlying
Higgs model, but their knowledge is indispensable to probe the existence of
additional sources of strong quantum effects beyond the SM.  Here we will
concentrate on the SUSY-QCD quantum effects mediated by squarks and gluinos
and shall compare them with the standard QCD corrections. A few remarks on the
remaining MSSM corrections will be made later on.

The relevant SUSY-QCD one-loop decay diagrams constructed from gluinos and
squarks (stop and sbottom species) are seen in Fig.1a.  In Fig.1b we sketch
one possible mechanism that could be used at the LHC to annually produce $\sim
10^4-10^6$ charged Higgs particles of a mass comprised between a few hundred
$GeV$ up to about $1\,TeV$ for a luminosity of ${\cal L}\sim
10^{34}\,cm^{-2}\,s^{-1}$\,\cite{Denegri}.  There are both the $t\,\bar{b}$-
and $b\,\bar{t}$-fusion mechanisms, which are to be treated together with the
processes $g\,\bar{b}\rightarrow H^+\,\bar{t}$ and $g\,b\rightarrow H^-\,t$ as
carefully explained in\,\cite{Soper}.  Trigging on a top quark in association
is very useful to avoid the signal being swamped by huge backgrounds.  These
mechanisms are primarily initiated by two highly energetic gluons, and use as
a production vertex the very same decay vertex that we wish to study. We point
out, in passing, that these diagrams also contribute to the cross-section for
single top-quark production, whose measurement is one of the main goals at the
next Tevatron run (Run II).  It is not our aim to give the complete list of
diagrams, but it is clear that these production mechanisms could be rather
efficient in the colliders, for the $H^{+}\,t\,\bar{b}$-vertex can be strongly
enhanced and result in a very distinctive phenomenology as compared to the
experimental expectations for the SM Higgs production, typically through
(one-loop) $g\,g$-fusion \,\cite{Denegri}. Most important, in the MSSM the
$H^{+}\,t\,\bar{b}$-vertex can receive significant corrections (in some cases
of order $50\%$) which could play a decisive role to disentangle whether a
charged Higgs hypothetically produced in a hadron collider is supersymmetric
or not.

To evaluate the relevant corrections to $\Gamma\equiv\Gamma (H^{+}\rightarrow
t\, \bar{b})$ in the MSSM, we shall adopt the on-shell renormalization
scheme\,\cite{BSH} where the fine structure constant, $\alpha$, and the masses
of the gauge bosons, fermions and scalars are the renormalized parameters:
$(\alpha, M_W, M_Z, M_H, m_f, M_{SUSY},...)$.  The interaction Lagrangian
describing the $H^{+}\,t\,\bar{b}$-vertex in the MSSM reads as follows: \beq
{\cal L}_{Htb}={g\,V_{tb}\over\sqrt{2}M_W}\,H^+\,\bar{t}\, [m_t\cot\beta\,P_L
+ m_b\tan\beta\,P_R]\,b+{\rm h.c.}\,,
\label{eq:LtbH}
\eeq where $P_{L,R}=1/2(1\mp\gamma_5)$ are the chiral projector operators,
$\tan\beta$ is the ratio between the vacuum expectation values of the two
Higgs doublets of the MSSM\,\cite{MSSM} and $V_{tb}$ is the corresponding CKM
matrix element--henceforth we set $V_{tb}=1$.

The on-shell renormalized vertex $H^+\,t\,\bar{b}$ in Fig.1a is derived
following a straightforward generalization of standard procedures in the
SM\,\cite{BSH} and can be parametrized in terms of two form factors $H_L$,
$H_R$ and the corresponding mass and wave-function renormalization
counterterms $\delta m_f$ and $\delta Z_{L,R}^f$ associated to the external
quark lines in Fig.1a: \beq i\,G={i\,g\over\sqrt{2}\,M_W}
\,\left[m_t\,\cot\beta\,(1+G_L)\,P_L + m_b\,\tan\beta\,(1+G_R)\,P_R\right]\,,
\label{eq:AtbH}
\eeq with \beqn G_L & = & H_L+{\delta m_t\over m_t} +\frac{1}{2}\,\delta
Z_L^b+\frac{1}{2}\,\delta Z_R^t \,,\nonumber\\ G_R &=& H_R+{\delta m_b\over
m_b} +\frac{1}{2}\,\delta Z_L^t+\frac{1}{2}\,\delta Z_R^b\,.
\label{eq:GLGR}
\eeqn The counterterms are computed in the on-shell scheme from the SUSY-QCD
interaction Lagrangian involving squarks and gluinos\,\cite{MSSM}.  The
explicit results, which we refrain from repeating here, are common to those in
Ref.\cite{GJS}.  As for the vertex form factors, they depend not only on pure
strong supersymmetric interactions but also on the Higgs-stop-sbottom semiweak
interaction Lagrangian\,\cite{MSSM}.  One finds (summation is understood over
indices $a,b=1,2$): \beqn H_L&=&8\pi\alpha _s\,i C_F\frac{G_{ab}^*}{m_t
\cot\beta} [R^{(t)}_{1 b}R^{(b)*}_{1 a}(\cuu-\cud)m_t+ R^{(t)}_{2
b}R^{(b)*}_{2 a}\cud m_b+ R^{(t)}_{2 b}R^{(b*)}_{1 a}\cz \mg]\,,\nonumber\\
H_R&=&8\pi\alpha _s\,i C_F\frac{G_{ab}^*}{m_b \tan\beta} [R^{(t)}_{2
b}R^{(b)*}_{2 a}(\cuu-\cud)m_t+ R^{(t)}_{1 b}R^{(b)*}_{1 a}\cud m_b+
R^{(t)}_{1 b}R^{(b)*}_{2 a}\cz \mg] \,,\nonumber\\
\label{eq:HLHR}
\eeqn where
$C_{...}=C_{...}(p,p^{\prime},m_{\tilde{g}},m_{\tilde{t}_b},m_{\tilde{b}_a})$
are standard three-point functions \cite{GJSH} also carrying indices $a,b$
summed over; $C_F=(N_C^2-1)/2N_C=4/3$ is a colour factor obtained after
summation over colour indices, and \beq G_{ab}=R^{(b)*}_{1 a}R^{(t)}_{1 b}
g_{LL}+ R^{(b)*}_{2 a}R^{(t)}_{2 b}g_{RR}+ R^{(b)*}_{2 a}R^{(t)}_{1 b}g_{LR}+
R^{(b)*}_{1 a}R^{(t)}_{2 b} g_{RL}\,, \eeq with \beqn g_{LL}&=&M_W^2\sin
2\beta -(m_t^2\cot\beta+m_b^2\tan\beta)\,,\nonumber\\
g_{RR}&=&-m_tm_b(\tan\beta+\cot\beta)\,,\nonumber\\
g_{LR}&=&-m_b(\mu+A_b\tan\beta)\,,\nonumber\\
g_{RL}&=&-m_t(\mu+A_t\cot\beta)\,.
\label{eq:GLR}
\eeqn The $2\times 2$ rotation matrices \beq R^{(q)} =\left(\begin{array}{cc}
\cos{\theta_q} & -\sin{\theta_q} \\ \sin{\theta_q} & \cos{\theta_q}
\end{array} \right)\;\;\;\;\;\;
(q=t, b)\,,
\label{eq:rotation}
\eeq plaguing the previous formulae relate the weak-eigenstate squarks
$\tilde{q'}_a=\{\tilde{q}_L,\tilde{q}_R\}$ to the mass-eigenstates
$\tilde{q}_a=\{\tilde{q}_1,\tilde{q}_2\}$ as follows: $\tilde{q'}_a=\sum_{b}
R_{ab}^{(q)}\tilde{q}_b$.  Therefore the $R^{(q)}$ diagonalize the
corresponding stop and sbottom mass matrices, whose standard form is
well-known\,\cite{MSSM} but we will quote here explicitly for convenience:
\begin{equation}
{\cal M}_{\tilde{t}}^2 =\left(\begin{array}{cc}
M_{\tilde{t}_L}^2+m_t^2+\cos{2\beta}({1\over 2}- {2\over 3}\,s_W^2)\,M_Z^2 &
m_t\, M_{LR}^t\\ m_t\, M_{LR}^t & M_{\tilde{t}_R}^2+m_t^2+{2\over
3}\,\cos{2\beta}\,s_W^2\,M_Z^2\,.
\end{array} \right)\,,
\label{eq:stopmatrix}
\end{equation}
\begin{equation}
{\cal M}_{\tilde{b}}^2 =\left(\begin{array}{cc}
M_{\tilde{b}_L}^2+m_b^2+\cos{2\beta}(-{1\over 2}+ {1\over 3}\,s_W^2)\,M_Z^2 &
m_b\, M_{LR}^b\\ m_b\, M_{LR}^b & M_{\tilde{b}_R}^2+m_b^2-{1\over
3}\,\cos{2\beta}\,s_W^2\,M_Z^2\,,
\end{array} \right)\,,
\label{eq:sbottommatrix}
\end{equation}
with \beq M_{LR}^t=A_t-\mu\cot\beta\,, \ \ \ \ M_{LR}^b=A_b-\mu\tan\beta\,.
\label{eq:MLRtb}
\eeq From the renormalized amplitude (\ref{eq:AtbH}) the width of
$H^+\rightarrow t\,\bar{b}$, including the one-loop SUSY-QCD corrections, is
the following: \beq \Gamma = \Gamma_0 \left\{1+\frac{U_L}{D}\,[2\,Re(G_L)]+
\frac{U_R}{D}\,[2\,Re(G_R)]+\frac{U_{LR}}{D}\, [2\,Re(G_L+G_R)]\right\}\,,
\label{eq:1Lwidth}
\eeq where the lowest-order result is \beq \Gamma_0=\left({N_C\,G_F\over
4\pi\sqrt{2}}\right){D\over\mh}\, \lambda^{1/2} (1,
{m_t^2\over\mhs},{m_b^2\over\mhs})\,,
\label{tree}
\eeq with $\lambda (1, x^2, y^2) = [1-(x+y)^2][1-(x-y)^2]$, and \beqn D &=&
(\mhs-m_t^2-m_b^2)\,(m_t^2\cot^2\beta+m_b^2\tan^2\beta)
-4m_t^2m_b^2\,,\nonumber\\ U_L & = &
(\mhs-m_t^2-m_b^2)\,m_t^2\cot^2\beta\,,\nonumber\\ U_R & = &
(\mhs-m_t^2-m_b^2)\,m_b^2\tan^2\beta\,,\nonumber\\ U_{LR} & = &
-2m_t^2m_b^2\,.  \eeqn The numerical analysis of the strong supersymmetric
corrections to $\Gamma (H^+\rightarrow t\,\bar{b})$ is presented in
Figs.2-5. In all figures where the Higgs mass is fixed, we take $M_{H^+}=
250\,GeV$, and we define $\alpha_s=\alpha_s(M_{H^+})$ by means of the
(one-loop) expression \beq \alpha_s(M_{H^+})={6\,\pi \over
(33-2\,n_f)\log\,({M_{H^+}/\Lambda_{n_f}})}\ , \eeq normalized as
$\alpha_s(M_Z)\simeq 0.12$, where $n_f$ is the number of quark flavors with
threshold below the Higgs boson mass $M_{H^+}$.

We treat the sbottom mass matrix (\ref{eq:sbottommatrix}) assuming a non-zero
mixing parameter $M_{LR}^b$. Once $\mu$ and $\tan\beta$ are given, we use
$A_b$ as one of the inputs.  However, for the sake of simplicity we assume
that $\theta_b=\pi/4$, so that the two diagonal entries are equal to a common
value.  We denote by $m_{\tilde{b}_1}$ the lightest sbottom mass-eigenvalue
and take it as the remaining input; typically, it is bound to satisfy
$m_{\tilde{b}_1}\stackM 100-150\,GeV$ from collider data, but the limits are
not so stringent as those from LEP, namely $m_{\tilde{b}_1}\stackM 65\,GeV$.
As for the stop mass matrix, we choose our inputs as follows. In this case the
lightest stop mass $m_{\tilde{t}_1}$ is strictly limited only by the LEP 1.5
bound $m_{\tilde{t}_1}\stackM 65\,GeV$, and we take it as one of the inputs.
Thus, upon fixing $\mu$, $\tan\beta$ and taking into account that
$SU(2)_L$-gauge invariance requires $M_{\tilde{t}_L}=M_{\tilde{b}_L}$, it
follows that the stop mass matrix depends on one remaining parameter which can
be taken as $A_t$.

To start with the numerical analysis, we study the dependence of the SUSY-QCD
effects on the crucial parameter $\tan\beta$.  In Fig.2a we plot the SUSY-QCD
corrected width, eq.(\ref{eq:1Lwidth}), versus $\tan\beta$, for fixed values
of the other parameters.  For completeness, we have included in this figure
the partial widths of the alternative decays
$H^+\rightarrow\tau^+\,\nu_{\tau}$ and $H^+\rightarrow W^+\,h^0$, which are
obviously free of ${\cal O}(\alpha_s)$ QCD corrections.  (To avoid cluttering,
we have not included $H^+\rightarrow c\,\bar{s}$; it is overwhelmed by the
$\tau$-lepton mode as soon as $\tan\beta\stackM 2$.)  It is patent from Fig.2a
that, for charged Higgs masses above the $t\,\bar{b}$ threshold, the decay
$H^+\rightarrow t\,\bar{b}$ is dominant.  Only for very large $\tan\beta$
($>30$) and for sufficiently big and positive $\mu$ ($\mu>100\,GeV$) --hence
outside Region II -- the negative corrections to $H^+\rightarrow t\,\bar{b}$
are huge enough to drive its partial width down to the level of
$H^+\rightarrow\tau^+\,\nu_{\tau}$.  Therefore, within the limits of Region
II, the top quark decay of the charged Higgs is, by far, the most relevant
decay mode to look at.

In considering the various parameter dependences, we present the results of
our analysis in terms of the quantity
\begin{equation}
\delta_{\tilde{g}}={\Gamma-\Gamma_0\over \Gamma_0}\,,
\label{eq:deltag}
\end{equation}
which gives the relative correction with respect to the tree-level width.  At
large $\tan\beta$, the role played by the bottom quark mass becomes very
important.  Indeed, in Fig.2b we confirm that the external self-energies
(basically the one from the $b$-line) give the bulk of the corrections
displayed in Fig.2a, whereas the (finite) vertex effect is comparatively much
smaller and its yield becomes rapidly saturated.  The existence of potentially
large SUSY-QCD corrections in the decay $H^+\rightarrow t\,\bar{b}$ could be
foreseen from the work of Ref.\,\cite{SO10}, where it is shown that the
SUSY-QCD bottom mass corrections are proportional to $\mu\tan\beta$ (for
fixed, nonvanishing, $\mu$ and large $\tan\beta$).  In our case, these
corrections are fed into the counterterm $\delta m_b/m_b$ on
eq.(\ref{eq:GLGR}) and, when viewed in terms of diagrams of the
electroweak-eigenstate basis, they appear as finite contributions proportional
to $M_{LR}^b$ generated from squark-gluino loops\footnote{In the absence of
sbottom mixing, i.e. $M_{LR}^b=0$, the large contribution $\delta m_b/m_b\sim
-\mu\tan\beta$ is no longer possible, but then the vertex correction does
precisely inherits this behaviour and compensates for it. However, the
condition $M_{LR}^b=0$, combined with a large value of $\tan\beta$, leads to
an scenario characterized by a value of $A_b$ which overshoots the natural
range expected for this parameter\,\cite{Referee}.}.

From Fig.3a we read off the incidence of the parameter $\mu$ on
$\delta_{\tilde{g}}$ for various $\tan\beta$.  We see that the SUSY-QCD
correction is extremely sensitive to $\mu$ both on its value and on its
sign. For this reason we have explored a moderate range of $\mu$ values.  It
turns out that the sign of the SUSY-QCD correction is basically opposite to
the sign of $\mu$, and the respective corrections for $+\mu$ and for $-\mu$
take on approximately the same absolute value.

Worth noticing is also the dependence on the gluino mass (Fig.3b).  The
various steep falls in that figure are associated to the presence of threshold
effects occurring at points satisfying $m_{\tilde{g}}+m_{\tilde{t}_1}\simeq
m_t$.  An analogous situation was observed in Ref.\cite{GJSH} for the SUSY
corrections to the standard top-quark decay. Away from the threshold points,
the behaviour of $\delta_{\tilde{g}}$ is smooth and perfectly consistent with
perturbation theory.  On the other hand, the sensitivity on $m_{\tilde{t}_1}$
is not dramatic, as can also be appraised in Fig.3b. This is to be expected
from the fact mentioned above that it is the bottom (not the top) self-energy
that dominates the corrections.  Virtually for any $m_{\tilde{t}_1}$, there
emerges an important correction which raises a long way with the gluino mass
before bending --very gently -- into the decoupling regime, as we have
checked.  The fact that the decoupling rate of the gluinos appears to be so
slow has an obvious phenomenological interest.
  
In Fig.4a we analyze in detail the SUSY-QCD correction as a function of the
sbottom masses.  We find convenient to plot $\delta_{\tilde{g}}$ versus
$m_{\tilde{b}_1}$, for fixed values of the other parameters and for various
$\tan\beta$.  We see that even for $m_{\tilde{b}_1}$ exceeding $250\,GeV$, and
$\tan\beta>10$, the correction remains sizeable ($\delta_{\tilde{g}}>10\%$).
As another feature, in Fig.4b we realize that $\delta_{\tilde{g}}$ is not
symmetric with respect to the sign of $A_b$.  Once the sign $\mu<0$ is chosen,
the correction is larger for negative values of $A_b$ than for positive
values. We have erred on the conservative side by choosing $A_b=+300\,GeV$
wherever this parameter is fixed.
 
In Fig.5 we compare the one-loop SUSY-QCD corrections with the standard ${\cal
O}(\alpha_s)$ QCD corrections as a function of $M_{H^+}$ and different values
of $\tan\beta$.  For the latter corrections we use the full analytical
formulae of Ref.\cite{Gambino}\footnote{We have corrected several misprints on
eq.(5.2) of Ref.\cite{Gambino}.}.  In the limit $M_{H^+}\gg m_t$, the standard
QCD correction boils down to the simple expression \beq
\delta_{g}={\Gamma_{QCD}-\Gamma_0\over \Gamma_0} =\left({C_F\,\alpha_s\over
2\,\pi}\right)\, {m_t^2\,\cot^2\beta\,\left({9\over 2}-6\,\log{M_{H^+}\over
m_t}\right)+ m_b^2\,\tan^2\beta\left({9\over 2}-6\,\log{M_{H^+}\over
m_b}\right) \over m_t^2\,\cot^2\beta+m_b^2\,\tan^2\beta}\,.
\label{eq:approx}
\eeq This formula is very convenient to understand the asymptotic behaviour.
However, as we have checked, it is inaccurate for the present range of values
of $m_t$ unless $M_{H^+}$ is extremely large (beyond $1\,TeV$).

Remarkably, we recognize from Fig.5 that the supersymmetric QCD effects
($\delta_{\tilde{g}}$) can be comparable or even larger than the conventional
QCD corrections ($\delta_g$).  For a given $\tan\beta$, the relative size of
the SUSY-QCD effects versus the standard QCD effects depends on the value of
$M_{H^+}$. Notwithstanding, it is clear that $\delta_{\tilde{g}}$ remains
fairly insensitive to $M_{H^+}$.

Some discussion on previous work is in order.  A first study of the SUSY-QCD
corrections to the hadronic width of a charged Higgs boson is performed in
Refs.\cite{LiYangKoenig}.  However, these references either use a too
simplified set of assumptions on the spectrum of gluinos and squarks (e.g. all
squarks are assumed unmixed and degenerate) and/or $m_b=0$ is assumed, so that
the local behaviours of $\delta_{\tilde{g}}$ and the crucial effect from
finite bottom mass corrections at large $\tan\beta$ are fully unnoticed.

Although we have concentrated on the computation of the QCD corrections, it is
legitimate to worry about the larger and far more complex body of electroweak
quantum effects, especially those coming from possible enhanced Yukawa
couplings.  A full analysis is under way and will be presented
elsewhere\,\cite{CGGJS}, but a few comments may be necessary here.  In fact,
Yukawa couplings can also give contributions to $\delta m_b/m_b$ proportional
to $\tan\beta$ \,\cite{SO10}, but their effect is in general smaller than in
the SUSY-QCD case. For example, if there is no large hierarchy between the
sparticle masses, the ratio between the SUSY-QCD and the Yukawa coupling
contributions to $\delta m_b/m_b$ can be estimated (at high $\tan\beta$) as
$4\,m_{\tilde{g}}/A_t$ times a slowly varying function of the masses of order
$1$ \,\cite{SO10}, where the (approximate) proportionality to the gluino mass
reflects the aforementioned very slowly decoupling rate of the latter.  In
view of the present bounds on the gluino mass, and since $A_t$ (as well as
$A_b$) cannot increase arbitrarily, we expect that the SUSY-QCD effects can be
dominant, and even overwhelming for sufficiently heavy gluinos.
Notwithstanding, the sole estimation in terms of $\delta m_b/m_b$ may be
insufficient, for there are also plenty of additional electroweak vertex
contributions both from the Higgs sector and from the
stop-sbottom/chargino-neutralino sector where those Yukawa couplings are also
involved. Hence, in contrast to the SUSY-QCD case, it is not obvious a priori
what is the net outcome of the leading electroweak contributions.  A partial
evaluation of the Yukawa coupling effects from supersymmetric Higgs bosons has
been made in Ref.\cite{YangLiHu} and entail only a few percent change in the
partial width.  Even though a complete calculation of the Higgs effects -- not
to mention those associated to the full plethora of supersymmetric particles
-- is not yet available in the literature, a preliminary analysis made by the
authors\,\cite{CGGJS} leads to the conclusion that the over-all leading
electroweak supersymmetric effects, though not negligible, remain
comparatively small and do not drastically alter the SUSY-QCD picture
presented here.

In conclusion, we have presented a fairly complete treatment of the SUSY-QCD
corrections to the partial width of the top quark decay of a charged Higgs
boson and have put forward plenty of evidence that they can be rather large
(typically between $10\%-50\%$), slowly decoupling and of both signs.
Consequently, they can either reinforce the conventional QCD corrections or
counterbalance them, and even reverse their sign; the QCD corrections would
then be found much ``larger'', ``missing'' or with the ``wrong'' sign,
respectively.  This should be helpful to differentiate $H^+$ from alternative
charged pseudoscalar decays leading to the same final states. Ultimately, the
precise knowledge of the quantum effects on $H^+\rightarrow t\,\bar{b}$ should
provide the characteristic features necessary to pin down the supersymmetric
nature of a heavy charged Higgs hypothetically discovered in the next
generation of experiments at the Tevatron and at the LHC.  As we have shown,
the mechanisms capable of producing a charged Higgs scalar in a hadron
collider, e.g.  $t\,\bar{b}$ and $b\,\bar{t}$ fusion, which can be greatly
enhanced at large $\tan\beta$, are very sensitive to potentially large
SUSY-QCD quantum effects.  If these effects are eventually found, they could
be the smoking gun needed to recognize that the produced $H^{\pm}$ in a hadron
collider is, truly, a SUSY Higgs.  After submiting this paper, we have noticed
the work of Ref.\cite{Bartl} dealing with the same subject.

{\bf Acknowledgements}:

\noindent
One of us (J.S.) has benefited from conversations with M. Mart\'\i nez and
F. Palla.
%on electroweak precision observables at LEP and on future Higgs
%physics at the hadron colliders.
We are very grateful to J.A. Coarasa for a careful reading of the manuscript
and for a numerical cross-check of the standard QCD contribution using an
independent numerical code based on Mathematica.  This work has been partially
supported by CICYT under project No. AEN95-0882.

%%%%%%%%%%%%%%%%%%%%%%%%%%%%%%%%%%%%%%%%%%%%%%%%%%%%%%%%%%%%%%%%%%%

\vspace{0.75cm}
\begin{center}
\begin{Large}
{\bf Figure Captions}
\end{Large}
\end{center}
\begin{itemize}
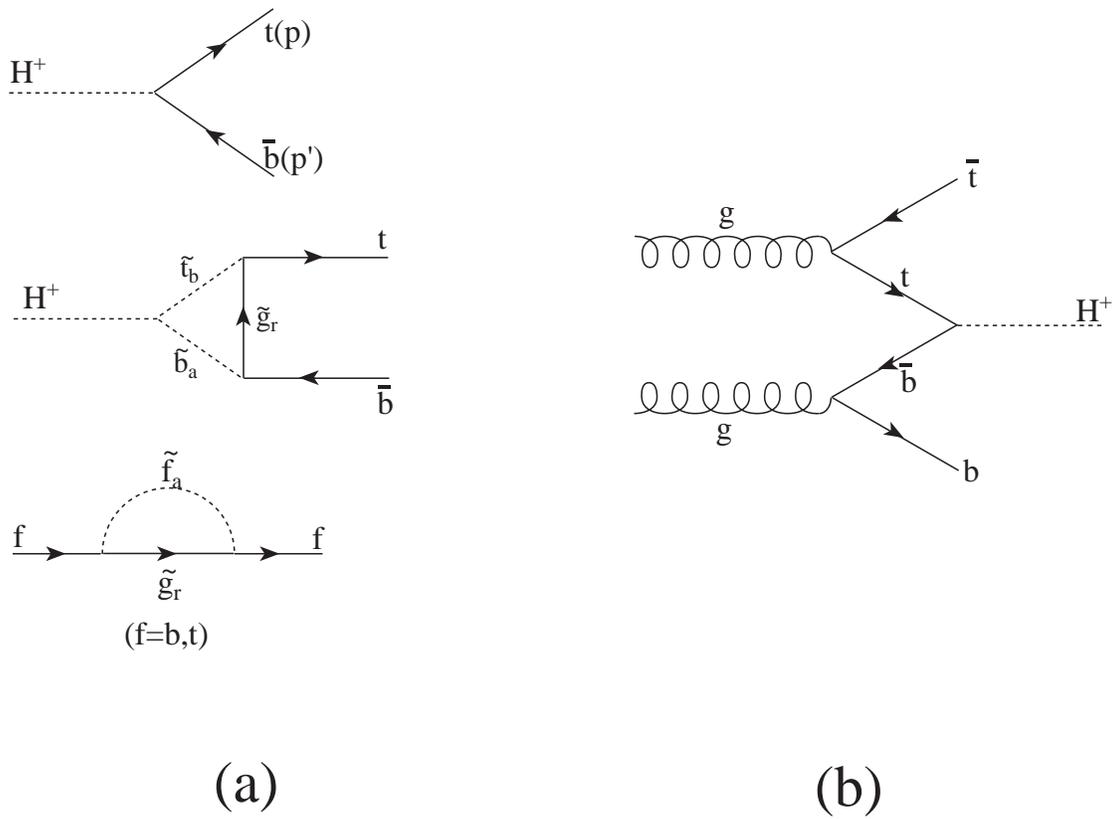
\item{\bf Fig.1} {\bf (a)} SUSY-QCD Feynman diagrams, up to one-loop order,
correcting the partial width of $H^{+}\rightarrow\,t\,\bar{b}$; {\bf (b)}
Typical charged-Higgs production mechanism at hadron colliders.

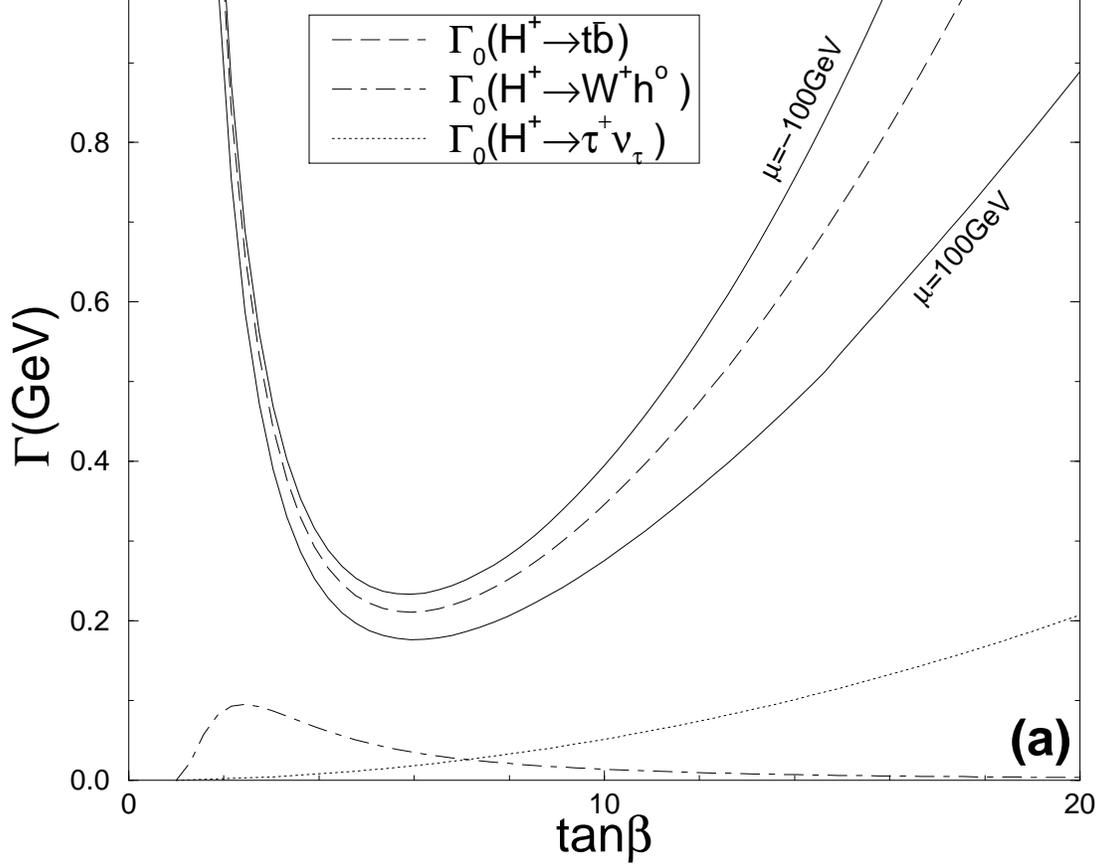
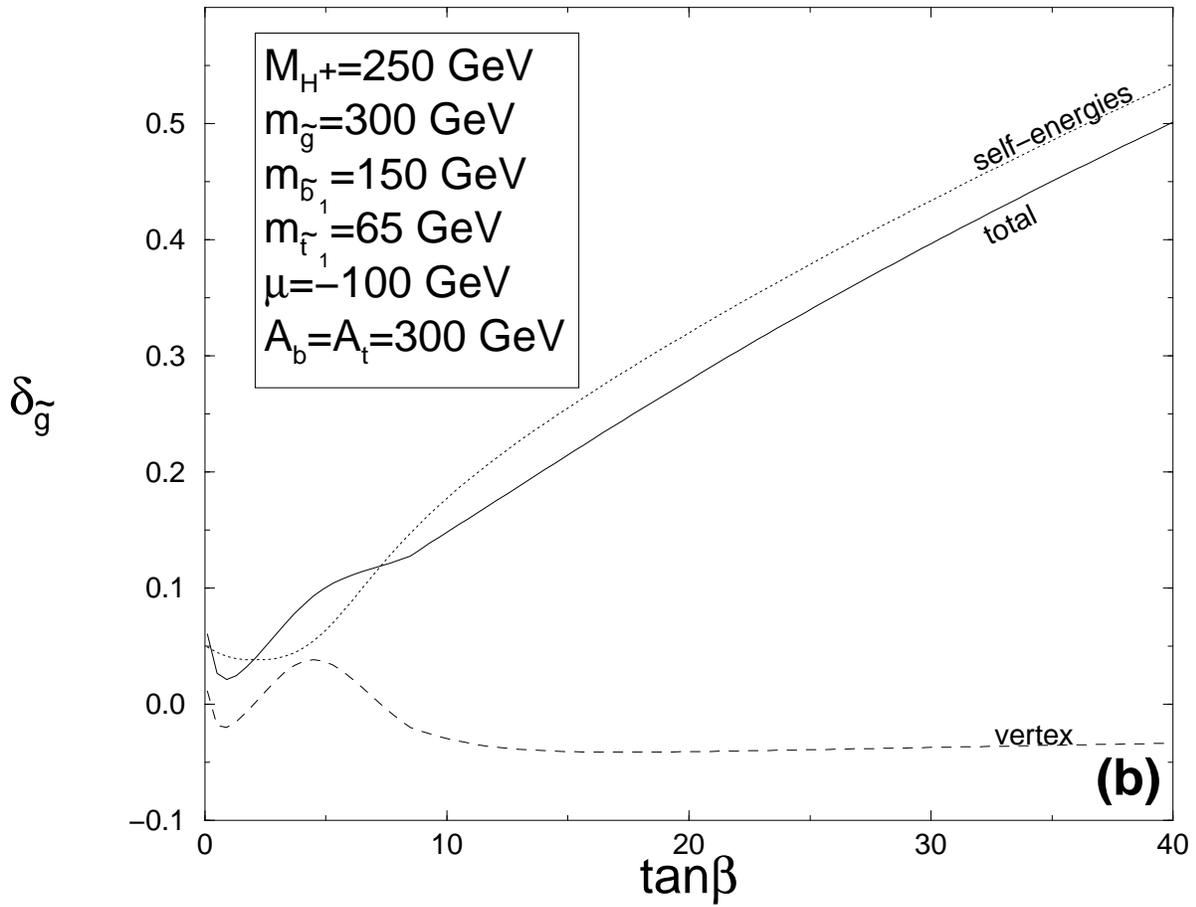
\item{\bf Fig.2} {\bf (a)} SUSY-QCD corrected $\Gamma( H^+\rightarrow
t\,\bar{b})$ -- eq.(\ref{eq:1Lwidth}) -- (in $GeV$) as a function of
$\tan\beta$, compared to the corresponding tree-level width, $\Gamma_0$.  Also
shown are the partial widths of the alternative decays
$H^+\rightarrow\tau^+\,\nu_{\tau}$ and $H^+\rightarrow W^+\,h^0$.  The
top-quark mass is $m_t=175\,GeV$; {\bf (b)} Comparative effects from the
(finite) vertex and self-energies as a function of $\tan\beta$. The fixed
parameters for (a) and (b) are given in the frame.

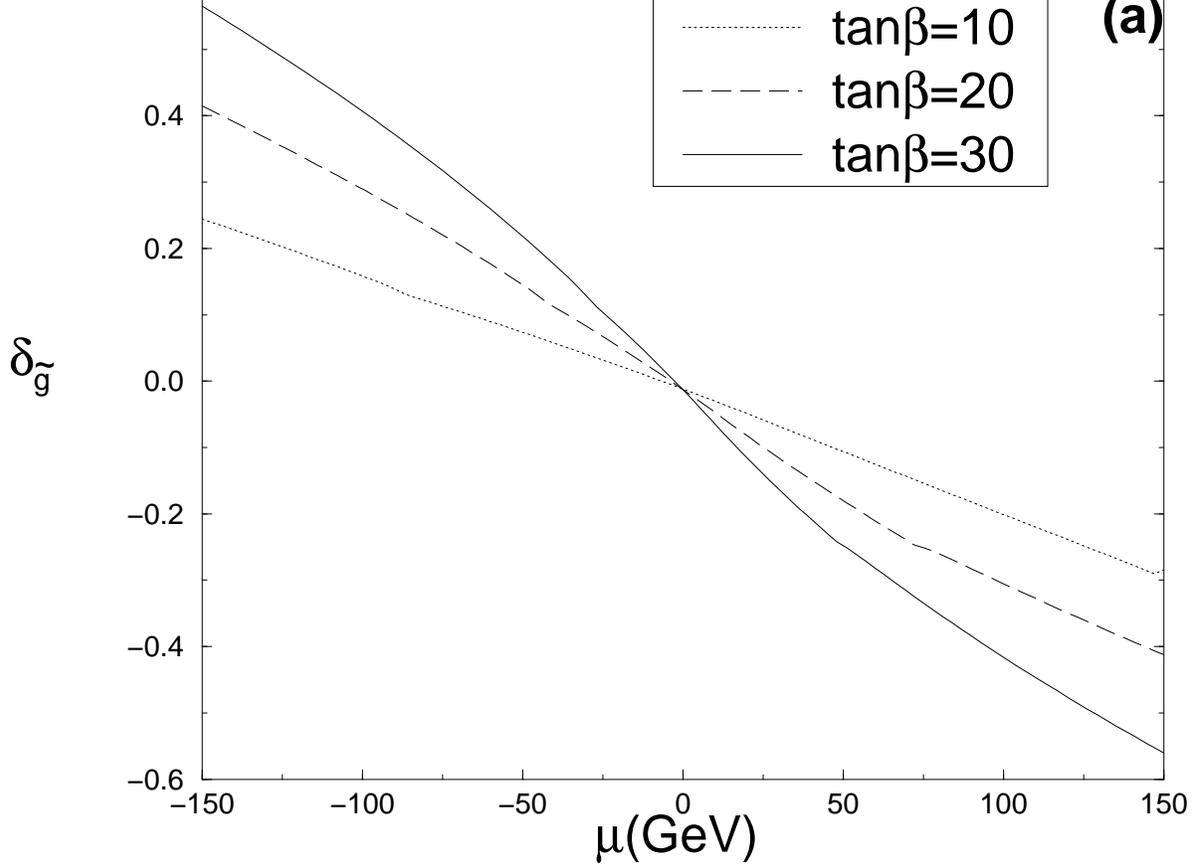
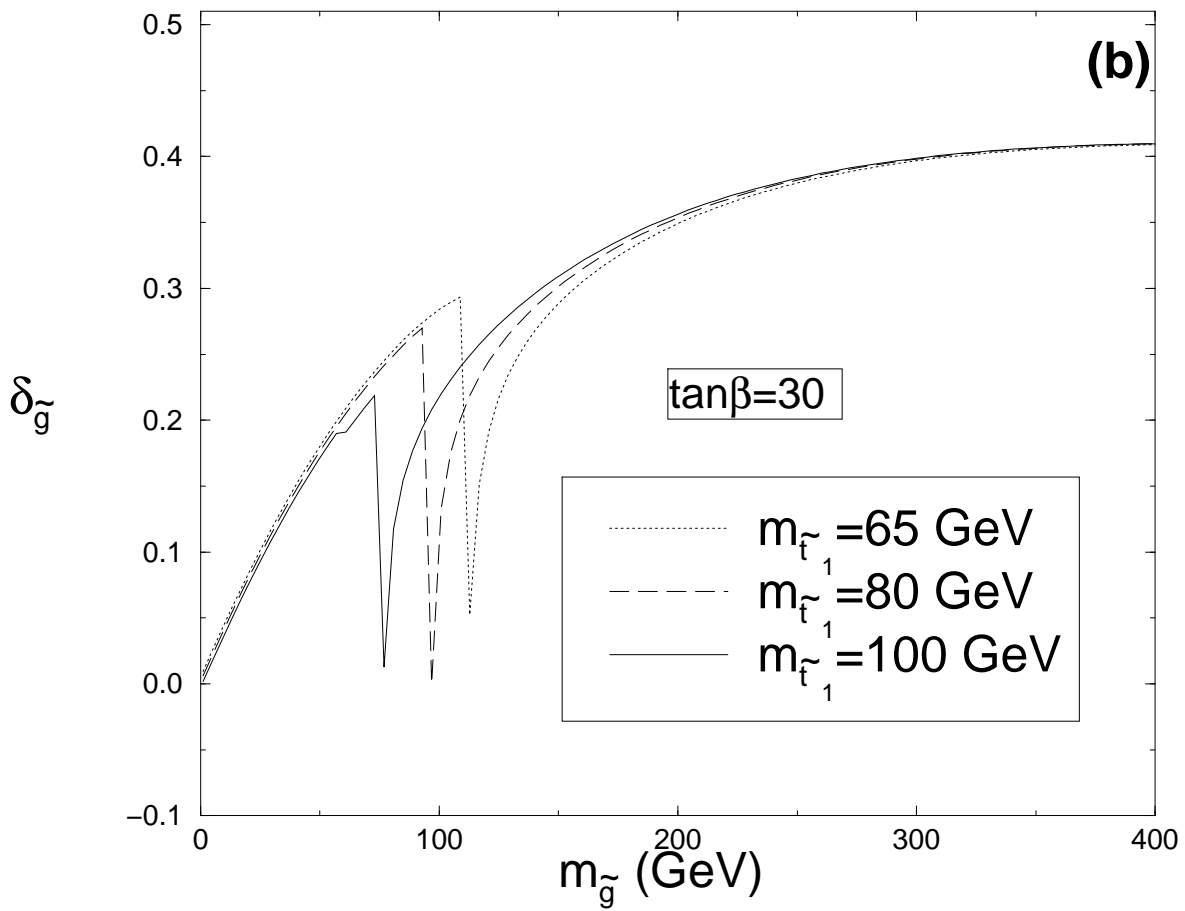
\item{\bf Fig.3} {\bf (a)} Dependence of the relative SUSY-QCD correction\,
$\delta_{\tilde{g}}$, eq.(\ref{eq:deltag}), upon the supersymmetric Higgs
mixing mass parameter, $\mu$; {\bf (b)} Evolution of the SUSY-QCD correction
in terms of the gluino mass for $m_{\tilde{t}_1}=65, 80,100\,GeV$,
$\tan\beta=30$. Rest of inputs as in Fig.2.

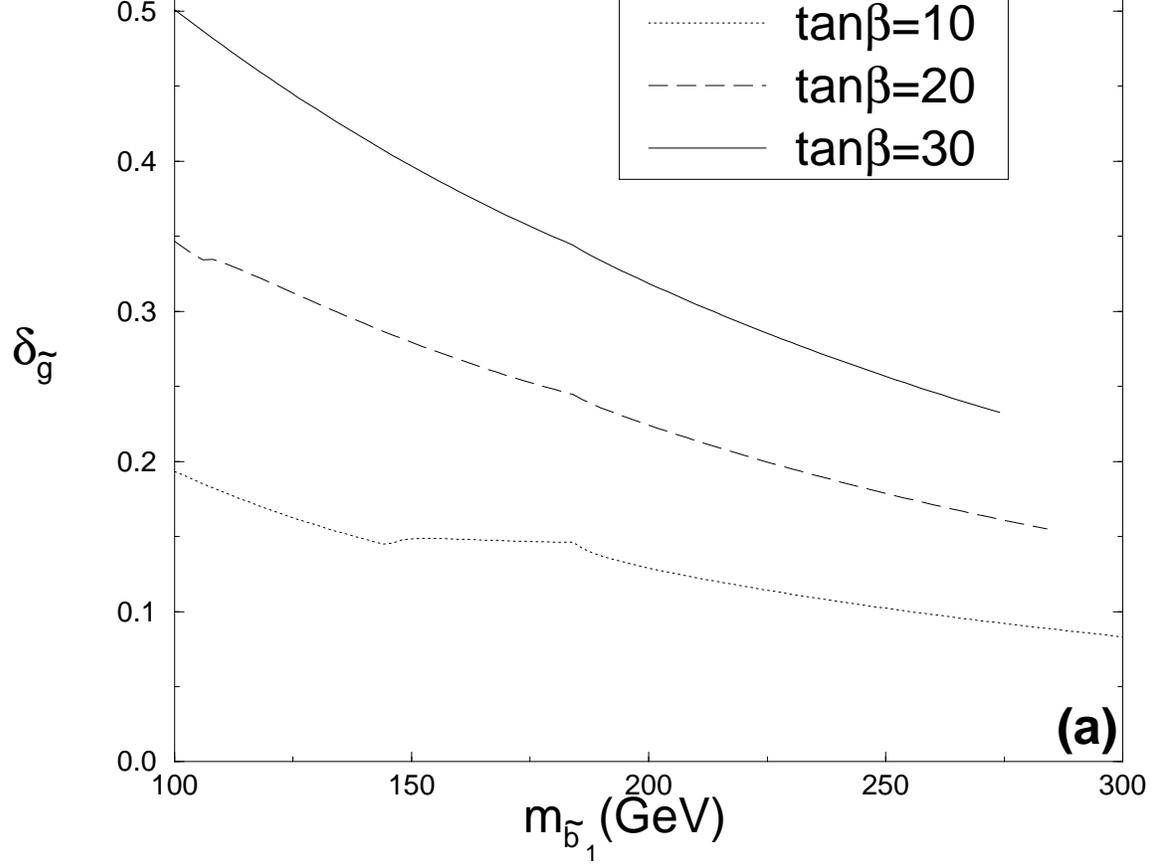
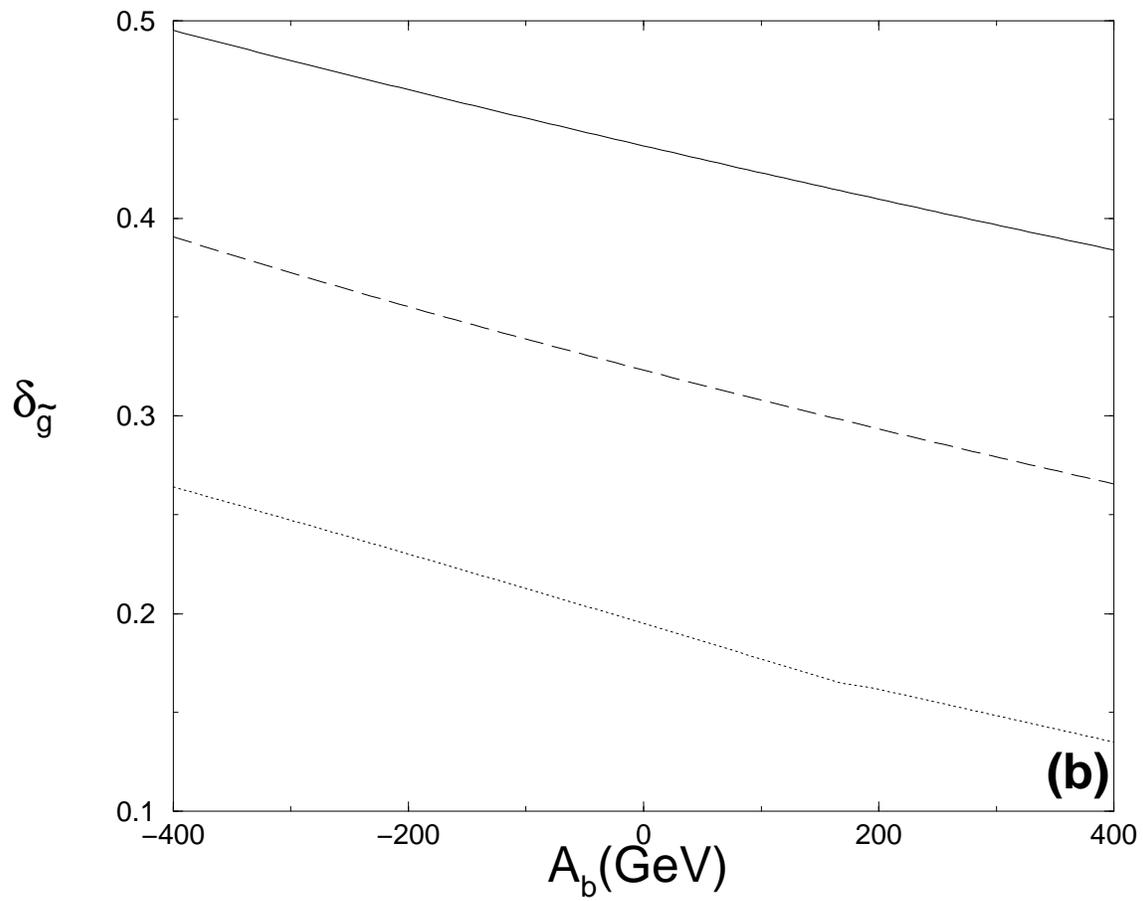
\item{\bf Fig.4} {\bf (a)} $\delta_{\tilde{g}}$ as a function of
$m_{\tilde{b}_1}$; {\bf (b)} $\delta_{\tilde{g}}$ as a function of $A_b$.  In
both cases $\tan\beta=10,20,30$ and the remaining inputs are as in Fig.2
    
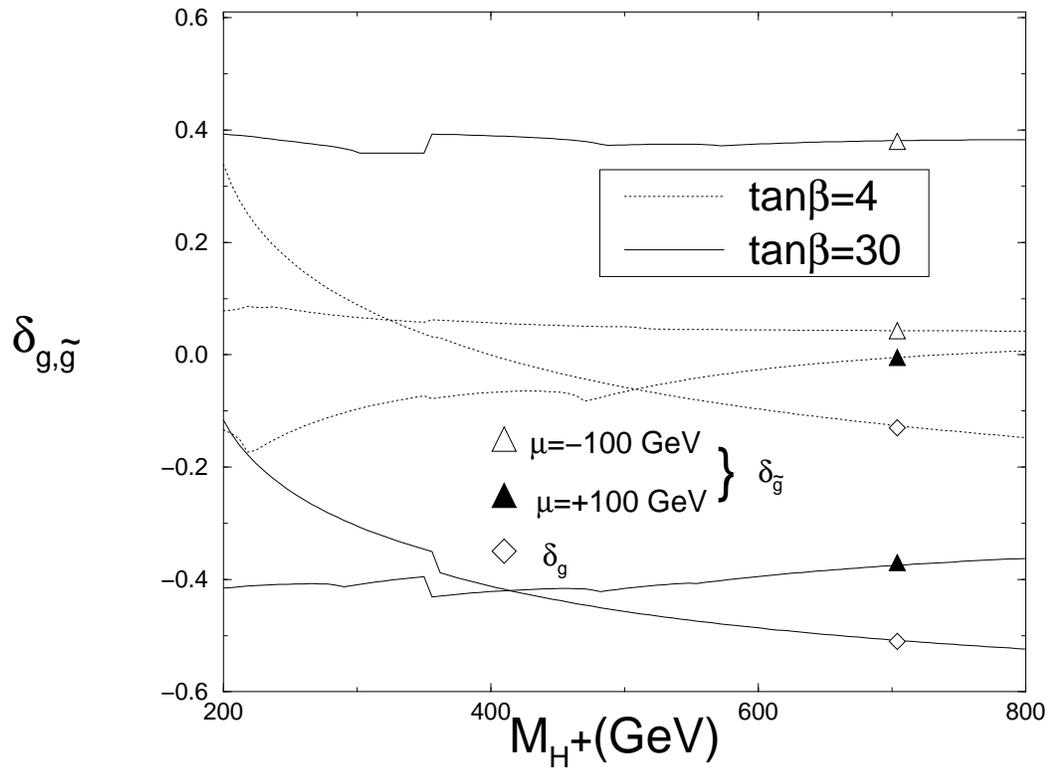
\item{\bf Fig.5} The ${\cal O}(\alpha_s)$ standard QCD correction, $\delta_g$,
compared to the SUSY-QCD correction, $\delta_{\tilde{g}}$, as a function of
$M_{H^+}$ and for two values of $\tan\beta$.  Rest of inputs as in Fig.2.
 
\end{itemize}

\end{document}